\documentclass[preprint2]{aastex}
\usepackage{color}

\shorttitle{Ruprecht 106}
\shortauthors{Villanova et al.}

\begin{document}

\title{Ruprecht 106: the first single population Globular
Cluster?}


\author{S. Villanova and D. Geisler}
\affil{Departamento de Astronom\'ia, Casilla 160, Universidad de Concepci\'on, Chile}
\email{svillanova@astro-udec.cl}

\author{G. Carraro}
\affil{European Southern Observatory, Alonso de Cordova 3107 Vitacura,
Santiago de Chile, Chile}

\author{C. Moni Bidin}
\affil{Instituto de Astronom\'{i}a, Universidad Cat\'{o}lica del Norte, Av. Angamos 0610, Antofagasta, Chile}

\and

\author{C. Mu\~noz}
\affil{Departamento de Astronom\'ia, Casilla 160, Universidad de Concepci\'on, Chile}




\begin{abstract}

All old Galactic Globular Clusters studied in detail to date host at least two
generations of stars,  where the second is formed from gas
polluted by processed material produced by massive stars of the first.
This process can happen if the initial mass of the cluster
exceeds a threshold above which ejecta are retained and a second
generation is formed.
A determination of this mass-threshold is mandatory in order to understand
how GCs form. 
We analyzed 9 RGB stars belonging to the cluster Ruprecht 106. Targets
were observed with the UVES@VLT2 spectrograph.
Spectra cover a wide range and allowed us to measure abundances for light
(O,Na,Mg,Al), $\alpha$ (Si,Ca,Ti) , iron-peak
(Sc,V,Cr,Mn,Fe,Co,Ni,Cu,Zn) and neutron-capture
(Y,Zr,Ba,La,Ce,Pr,Nd,Sm,Eu,Dy,Pb) elements.
Based on these abundances we show that Ruprecht 106 is the first
convincing example of a single population GC (i.e. a true simple stellar
population), although the sample is relatively small. This result is supported also by
an independent photometric test and by the HB morphology and the
dynamical state.
It is old ($\sim$12 Gyrs) and, at odds with other GCs, has no $\alpha$-enhancement. The
 material it formed from was contaminated by both s- and
r- process elements. The abundance pattern points toward an extragalactic origin.
Its present day mass (M=10$^{4.83}$ M$_{\odot}$) can be assumed as a strong lower limit for the
initial mass threshold below which no second generation is
formed. Clearly, its initial mass must have been significantly greater but we have no
current constraints on the amount of mass loss during its evolution.

\end{abstract}


\section{Introduction}

Recently Globular Clusters (GCs) in the Galaxy were discovered to have chemical inhomogeneities.  
More specifically, \citet{Ca09} showed that all GCs studied up to now have at least a
spread (or anticorrelation) in the content of their light-elements O and Na.
Indeed, they present a new definition of a GC as a cluster which exhibits such
an anticorrelation, with the implication that all globulars, at least those
above a certain mass limit, must posses this characteristic.
Many other light-elements such as C,N,Mg, and Al also show a spread or a (anti)correlation. 
The Na-O anticorrelation was found over the entire mass range observed,
from NGC~6838 (M=10$^{4.30}$ M$_{\odot}$) up to 47~Tuc (M=10$^{6.03}$
M$_{\odot}$). This spread is probably due to the early evolution of each
cluster, when a second generation of stars was born from gas polluted by
ejecta of evolved stars of the first generation (the so called
multiple-population phenomenon, \citealt{Kr94,Gr04,Gr12,Pi09,Pi12}). 
According to this model, stars of the first (older) generation were born with
normal He (Y$\sim$0.25) and a Na/O content similar to the field
stars in the Halo, while stars of the second (younger) generation, because of this
self-enrichment, are He/Na richer (Y$\geq$0.25, \citealt{Da08}) and O-poorer with respect to the
first generation. Several kinds of polluters have been proposed, including intermediate
mass AGB stars (M$\simeq$4-8 M$_{\odot}$, \citealt{Ve01}), fast rotating massive stars (M$>$20 M$_{\odot}$, \citealt {De07})
massive binaries stars (M$\simeq$20 M$_{\odot}$, \citealt{Mi09})  and novae \citep{Mac12}. In some cases also
SNeII may have been at work \citep{Ma09}.
The first requirement for this process is that the initial mass of
the cluster was high enough to retain both some primordial gas and the
ejecta. The higher the initial mass, the higher the mass of the gas and ejecta
that can be retained, and the more extended the abundance spread that is expected to
be observed nowadays \citep{De08}. Subsequently, all clusters are expected to
lose mass due to both internal and external processes.

The nature of the most effective polluter changes with cluster mass, because more
massive GCs can retain faster ejecta, including those from a SNe explosion \citep{Va08}. 
This is the case for M22\citep{Ma09}, $\omega$ Centauri \citep{Jo10}, and M54
\citep{Ca10},  where spread in the $\alpha$-element and iron content were found besides the usual
light-element spread, indicating pollution by SNeII. 

Lower mass GCs can retain only slow winds, such as those from massive main sequence stars
or intermediate mass AGB stars where only light-element variation is expected
(as in the case of the standard globular cluster M4, \citealt{Ma08}).

Below a certain mass threshold no ejecta are retained at all, so single
population GCs are expected \citep{Ca11}. Indeed, searches for light element
spreads in much lower mass open clusters have been negative \citep{De09}, with
the exception of NGC~6791 \citep{Ge12} that, however, may not be a genuine member
of this category \citep{Ca12}.

\citet{Ca09} showed that most of the stars currently found in a GC belong to the second
generation ($\sim$60$\div$80\%). This is at odds with theory, which says
that first generation stars must have been much more numerous than we
observe nowadays in order to produce enough ejecta to form the second. 
This contradiction can be partially explained if we assume that ejecta were
collected preferentially in
the center of the cluster due to the gravitational potential. Because of this
the second generation was formed in the center and was much less affected by
Galactic tidal disruption than the first, which lost most of its members \citep{Ca11}.

This scenario holds also for old and intermediate-age massive clusters in the
Large Magellanic Cloud (LMC, \citealt{Mu08,Mu09}). The former are generally
more massive than the latter and appear to mimic Galactic GCs in having an
extended Na/O anticorrelation, while the latter only show some spread in Na,
with the possible exception of a single object (see section 5) that
is doubtful. For this reason \citet{Mu09} reiterated that mass can
be the key factor in determining chemical inhomogeneities.

Recently \citet{Ca11} suggested the possibility of the existence of two types of
single population GCs. The first is represented by clusters initially not
massive enough to be able to retain primordial gas and ejecta from evolved stars and form
a second generation of stars. The second is represented by massive clusters 
that retained almost all the first generation and so only a small
fraction of the stars would belong to the second generation population.
Such clusters would preferentially be those that do not fill their tidal radius.
These second are only pseudo-single population clusters, because the second generation of stars is
just very small compared to the first generation, but still present.
\citet{Ca11} presented Palomar 3 as a probable example of the first type
because of the small color spread of its horizontal branch (HB), of only
$\sim$0.25 mag in B-V (see their Fig. 2).

A key question is if there is any relatively massive globular cluster that is
composed of only chemically homogeneous stars. Such a cluster should be sought
among the less massive globulars in the Galaxy.
For this reason we focused our investigation on Ruprecht 106 
(M$\sim$10$^{4.8}$ M$_{\odot}$, \citealt{Ma91}).
This cluster was first studied by \citet{Bu90} who suggested an age 4-5
Gyrs younger then the oldest halo GCs and a metallicity of [Fe/H]$\sim$-2.0
based on photometric indicators. \citet{Fr97} gave [Fe/H]$\sim$-1.6,
based on spectroscopic observations.
According to the most recent work by \citet{Do11} Ruprecht 106 is a
relatively old, metal-poor (11.5 Gyrs, [Fe/H]$\sim$-1.5) GC with a solar scaled
$\alpha$-element content.

We selected this cluster, apart from its low mass, for the small extension of its
HB. According to \citet{Da02}, the extension of the HB in a GC is proportional to
the amount of helium variation due to self-pollution among its stars. 
He-normal stars are located in the redder part of the HB, while He-rich stars
lie in the bluer part, as recently shown by \citet{Ma11} and \citet{Gr11,Gr12,Gr13}.

\citet{Ca11} suggest that the best candidates for single population GCs
are those with both stubby HBs for their metallicity and  with
M$<$10$^{4.8}$ M$_{\odot}$. Ruprecht 106 fulfills both of these
characteristics but was not investigated by them.

\begin{figure}[ht!]
\epsscale{1.00}
\plotone{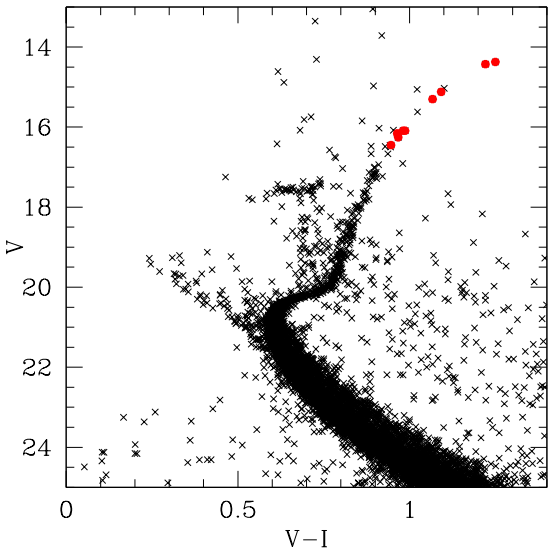}
\caption{The CMD of Ruprecht 106 with the observed RGB stars indicated as filled
  red circles \citep{Do11}.}
\label{f1}
\end{figure}

So an HB with a small extension may indicate no spread in
helium (or light-elements) and the absence of the self-pollution
phenomenon. In such a cluster only one
generation (the first) should be present. It should also have a
homogeneous content of all the other elements ($\alpha$, iron-peak, s and r).
The scenario is complicated by the fact that age and mass-loss are also involved.
In fact a young cluster (less then 10 Gyrs) could show no
color spread on the HB (that would appear entirely red) but still have a
variation in He just because HB stars do not have a low enough mass to be
located in the blue part (the younger the age, the higher the mean mass in the HB).
On the other hand a differential mass-loss along the HB can generate a
spread on the HB even if no He variation is present.
Ruprecht 106 is old and metal poor, so it should have a blue HB, or at least
some of the HB stars should be located on the blue HB, as in M4 or NGC~6752,
two clusters with an extended and well studied Na-O anticorrelation \citep{Ma08,Ca07}.
In spite of this, \citet{Do11} show a CMD (see Fig.~\ref{f1}) with only a red
and slightly extended HB. The color baseline of the HB is
$\sim$0.2 mag, less than Palomar 3, which was proposed by \citet{Ca11} as a
probable single population cluster.
For all these reasons we selected it as a good candidate single population GC.
In order to prove if this is true, we analized 
spectra for a statistically significant sample of RGB stars and measured their
chemical composition. If it indeed is a single population cluster we should observe a
homogeneous content of all the elements, within the observational errors.

In section 2 we describe data reduction and in section 3 the methodology we
used to obtain the chemical abundances. In section 4 we present our results
including a comparison with the literature. In section 5 and 6 we discuss Ruprecht 106
as a single population GC, and in section 7 we compare it with different
formation environments (Galactic and extragalactic). Finally in section 8 we give a
summary of our findings.

\section{Observations and data reduction}

Our dataset consists of high resolution spectra collected in 2002 and
downloaded from the Advanced Data Products ESO archive \footnote{http://archive.eso.org/eso/eso\_archive\_adp.html}.
The spectra were obtained with the UVES spectrograph, mounted at the VLT telescope.
A total of 10 stars were observed between V=14.4 and V=16.5,from the RGB-bump
up to the RGB-tip of the cluster (see Fig.~\ref{f1}). 
All stars were observed with the blue and red arms of the spectrograph, and
spectra cover a range of 3700--6800 \AA. The signal-to-noise (S/N) is between 50 and 70 at 6000 \AA.
 
Data were reduced using the dedicated pipeline (see
http://www.eso.org/sci/software/pipelines/).
Data reduction includes bias subtraction, flat-field correction, wavelength calibration,
sky subtraction, and spectral rectification.

Radial velocities were measured by the {\it fxcor} package in IRAF,
using a synthetic spectrum as a template. One star turned out to have a very
different radial velocity, and so was rejected as a non-member.
The mean heliocentric value for our member targets is -38.4$\pm$0.4 km/s.
\citet{DC92} gives -44$\pm$3 km/s.
There is a discrepancy between the two results; however, it is not dramatic
and due probably to the low resolution of \citet{DC92} spectra ($\Delta$$\lambda$=3.2 \AA).

Table~\ref{t1} lists the basic parameters of the member stars:
ID (from \citealt{Bu90}), J2000.0 coordinates (RA \& DEC), ACS@HST V and I magnitudes (Sarajedini, private
communication), heliocentric radial velocity (RV$_{\rm H}$),
T$_{\rm {eff}}$, log(g), micro-turbulence velocity (v$_{\rm t}$), and
metallicity ([Fe/H]). The determination of the atmospheric parameters is
discussed in the next section.

\begin{deluxetable}{lccccccccc}
\tablecolumns{10}
\tablewidth{0pc}
\tablecaption{Basic parameters of the observed stars.}
\tablehead{
\colhead{ID} & \colhead{RA(h:m:s)} & \colhead{DEC($^{o}$:':'')} & V(mag) & I(mag) & \colhead{RV$_{\rm H}$(km/s)} &
\colhead{T$_{\rm {eff}}$(K)} & log(g)(dex) & \colhead{v$_{\rm t}$(km/s)} & [Fe/H] 
}
\startdata
\hline
Ru$_{1445}$ & 12:38:39.2 &-51:08:41.4 &16.257 &15.290 &-40.26 &4580 &0.80 &1.52 &-1.52\\
Ru$_{1614}$ & 12:38:36.4 &-51:08:24.5 &14.373 &13.123 &-36.53 &4020 &0.00 &1.80 &-1.44\\ 
Ru$_{1863}$ & 12:38:34.9 &-51:08:03.8 &16.090 &15.109 &-37.79 &4600 &1.00 &1.50 &-1.49\\  
Ru$_{1951}$ & 12:38:35.2 &-51:07:51.6 &15.120 &14.028 &-36.51 &4380 &0.60 &1.60 &-1.39\\  
Ru$_{2004}$ & 12:38:40.7 &-51:07:46.1 &14.427 &13.206 &-39.12 &4140 &0.10 &1.74 &-1.46\\  
Ru$_{2032}$ & 12:38:48.8 &-51:07:43.2 &16.160 &15.196 &-38.25 &4570 &1.05 &1.44 &-1.49\\  
Ru$_{676}$  & 12:38:44.0 &-51:09:51.7 &16.091 &15.104 &-38.51 &4550 &0.85 &1.50 &-1.52\\  
Ru$_{801}$  & 12:38:36.2 &-51:09:37.6 &16.450 &15.504 &-39.56 &4600 &0.95 &1.51 &-1.51\\   
Ru$_{970}$  & 12:38:42.4 &-51:09:22.3 &15.301 &14.234 &-39.41 &4400 &0.55 &1.60 &-1.42\\  
\hline
\enddata
\label{t1}
\end{deluxetable}

\section{Abundance analysis}

The chemical abundances for Mg, Si, Ca, Ti, Cr, Fe, and Ni were
obtained using the equivalent widths (EQWs) method.
See \citet{Ma08} for a more detailed explanation of the method we
used to measure the EQWs.
For the other elements (O, Na, Al, Sc, V, Mn, Co, Cu, Zn, Y, Zr, Ba, La,
Ce, Pr, Nd, Sm, Eu, Dy, Pb), whose lines are affected by blending, we used the
spectrum-synthesis method. For this
purpose we calculated 5 synthetic spectra having different abundances,
and estimated the best-fitting value as the one that minimizes the
r.m.s. scatter.
Si presents few features in the spectrum, so in this case abundances derived from
the EQWs were cross-checked with the spectral synthesis method in order to
obtain more accurate measurements. 
Only lines not contaminated by telluric features were used. 

Initial atmospheric parameters were obtained in the following way. 
First, T$_{\rm eff}$ was derived from the V-I color using the relation of
\citet{Al99} and the reddening (E(B-V)=0.20) from \citet{Ha96}. 
Surface gravities (log(g)) were obtained from the canonical equation:
$$ \log\left(\frac{g}{g_{\odot}}\right) =
         \log\left(\frac{M}{M_{\odot}}\right)
         + 4 \log\left(\frac{T_{\rm{eff}}}{T_{\odot}}\right)
         - \log\left(\frac{L}{L_{\odot}}\right). $$
where the mass M/M$_{\odot}$ was assumed to be 0.8 M$_{\odot}$, and the
luminosity L/L$_{\odot}$ was obtained from the absolute magnitude M$_{\rm V}$
assuming an apparent distance modulus of (m-M)$_{\rm V}$=17.25 \citep{Ha96}. The
bolometric correction (BC) was derived by adopting the relation 
BC-T$_{\rm eff}$ from \citet{Al99}.
Finally, micro-turbulence velocity (v$_{\rm t}$) was obtained from the
relation of \citet{Ma08}.\\
These atmospheric parameters were considered as initial estimates and were refined during the
abundance analysis. As a first step, atmospheric models were calculated using ATLAS9 \citep{Ku70}
assuming the initial estimate of T$_{\rm eff}$, log(g),
and v$_{\rm t}$, and the [Fe/H] value from \citet{Ha96}([Fe/H]=-1.68).\\ 
Then T$_{\rm eff}$, v$_{\rm t}$, and log(g) were adjusted and new
atmospheric models calculated in an interactive way in order to remove trends in
Excitation Potential (E.P.) and equivalent width vs. abundance for 
T$_{\rm eff}$ and v$_{\rm t}$ respectively, and to satisfy the ionization
equilibrium for log(g). FeI and FeII were used for this latter purpose. 
The [Fe/H] value of the model was changed at each iteration according to the
output of the abundance analysis.
The Local Thermodynamic Equilibrium (LTE) program MOOG \citep{Sn73} was used
for the abundance analysis.
Na is known to be affected by departure from LTE, so we applied the
\citet{Ma00} NLTE correction to our Na abundances. Due to the small
T$_{\rm eff}$ range covered by our stars, we decided to apply a mean
correction of -0.20 dex to all the stars.

The linelist is that used in previous papers (e.g. \citealt{Vi11}),
so we refer to those articles for detailed discussion about this point.
The adopted solar abundances we used are reported in Tab.~\ref{t2} and agree well
with those given by \citet{Gr98}.

\begin{figure}
\epsscale{1.00}
\plotone{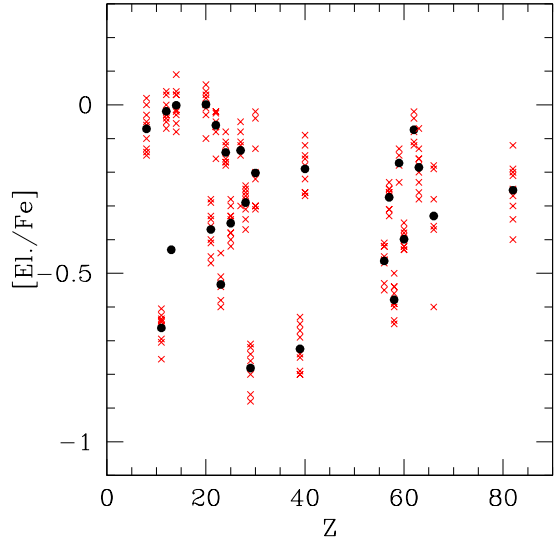}
\caption{Element abundance for the single stars (red crosses) and for the
cluster mean (black filled circles). See text for more details.}
\label{f2}
\end{figure}

Apart from elements already measured in our previous papers, here we added
Sc (5684 \AA\ line), V (6275 and 6285 \AA\ lines), Mn (5420 \AA\ line), Co
(5248 \AA\ line), Cu (5218 \AA\ line), Zn (4811 \AA\ line), La (5123 \AA\
line), Ce (5274 \AA\ line), Pr (4497 \AA\ line), Nd (5320 \AA\ line), Sm (4499
\AA\ line), Dy (5170 \AA\ line), Pb (4058 \AA\ line).
We could measure Al only for the star Ru$_{1614}$. For the other targets we
give upper limits. Element abundances are reported in Tab.~\ref{t2} and
Fig.~\ref{f2}.

A detailed internal error analysis was performed by varying T$_{\rm eff}$, log(g), [Fe/H], and
v$_{\rm t}$ and redetermining abundances of star \#Ru$_{1951}$, assumed to represent
the entire sample. Parameters were varied by $\Delta$T$_{\rm eff}$=+30 K,
$\Delta$log(g)=+0.09, $\Delta$[Fe/H]=+0.05 dex, and $\Delta$v$_{\rm t}$=+0.03
km/s. This estimation of the internal errors for atmospheric parameters was
performed as in \citet{Ma08}.
Results are shown in Tab.~\ref{t3}, including the error due to the noise
of the spectra. This error was obtained for elements whose abundance was
obtained by EQWs, as the average value of the errors of the mean given by
MOOG, and for elements whose abundance was obtained by spectrum-synthesis, as
the error given by the fitting procedure. $\sigma_{\rm tot}$ is the
squared sum of the single errors, while $\sigma_{\rm obs}$ is the mean
observed dispersion. The agreement between the two values is
well within 3$\sigma$ for all elements, indicating that there is no evidence for chemical
inhomogeneity for the 9 giants studied in Ruprecht 106. This is our principle finding.
Only for Ce we have a 3$\sigma$ difference, while for Sm $\sigma_{\rm tot}$ differs from
$\sigma_{\rm obs}$ by 6$\sigma$ and is larger. This suggests some
overestimation of the internal error.

\begin{deluxetable}{lccccccccccc}
\tablecolumns{13}
\tablewidth{0pc}
\tablecaption{Columns 2-10: abundances of the observed stars. Column 11: mean abundance for the cluster.
Column 12: abundances adopted for the Sun in this paper.
Abundances for the Sun are indicated as log$\epsilon$(El.)}
\tablehead{
El. & \colhead{Ru$_{1445}$}& \colhead{Ru$_{1614}$}& \colhead{Ru$_{1863}$}&
\colhead{Ru$_{1951}$}& \colhead{Ru$_{2004}$}& \colhead{Ru$_{2032}$}&  
\colhead{Ru$_{676}$}& \colhead{Ru$_{801}$}& \colhead{Ru$_{970}$}& Cluster & Sun
}
\startdata
\hline
${\rm [O/Fe]}$  &  -0.15 &  +0.02 &  -0.13 &  -0.06 &  -0.03 &  +0.00 &  -0.10 &  -0.05 &  -0.14 & -0.07$\pm$0.02& 8.83\\ 
${\rm [Na/Fe]}$ &  -0.41 &  -0.45 &  -0.44 &  -0.56 &  -0.50 &  -0.46 &  -0.43 &  -0.51 &  -0.44 & -0.46$\pm$0.02&  -  \\   
${\rm [Na/Fe]_{NLTE}}$ &  -0.61 &  -0.65 &  -0.64 &  -0.76 &  -0.70 &  -0.66 &  -0.63 &  -0.71 &  -0.64 & -0.66$\pm$0.02& 6.32\\
${\rm [Mg/Fe]}$ &  -0.07 &  -0.03 &  -0.04 &  +0.03 &  -0.05 &  -0.02 &  +0.00 &  +0.04 &  -0.03 & -0.02$\pm$0.01& 7.56\\  
${\rm [Al/Fe]}$ &$<$-0.31&  -0.43 &$<$-0.54&$<$-0.24&$<$-0.47&$<$-0.54&$<$-0.01&$<$+0.08&$<$-0.41& -0.43$\pm$0.10& 6.43\\   
${\rm [Si/Fe]}$ &  +0.09 &  +0.03 &  -0.02 &  -0.02 &  -0.06 &  +0.03 &  -0.03 &  +0.04 &  -0.08 & +0.00$\pm$0.02& 7.61\\    
${\rm [Ca/Fe]}$ &  +0.06 &  -0.10 &  +0.02 &  +0.00 &  -0.03 &  +0.00 &  +0.03 &  -0.01 &  +0.04 & +0.00$\pm$0.02& 6.39\\  
${\rm [Sc/Fe]}$ &  -0.45 &  -0.29 &  -0.41 &  -0.33 &  -0.28 &  -0.34 &  -0.40 &  -0.36 &  -0.47 & -0.37$\pm$0.03& 3.12\\ 
${\rm [Ti/Fe]}$ &  -0.07 &  -0.06 &  -0.07 &  -0.02 &  -0.02 &  -0.05 &  -0.08 &  -0.16 &  -0.03 & -0.06$\pm$0.01& 4.94\\
${\rm [V/Fe]}$  &  -0.44 &  -0.60 &  -0.54 &  -0.58 &  -0.51 &    -   &    -   &    -   &  -0.53 & -0.53$\pm$0.02& 4.00\\
${\rm [Cr/Fe]}$ &  -0.16 &  -0.14 &  -0.18 &  -0.08 &  -0.11 &  -0.17 &  -0.14 &  -0.17 &  -0.12 & -0.14$\pm$0.01& 5.63\\
${\rm [Mn/Fe]}$ &  -0.34 &  -0.38 &  -0.33 &  -0.33 &  -0.28 &  -0.38 &  -0.40 &  -0.42 &  -0.30 & -0.35$\pm$0.02& 5.37\\
${\rm [Fe/H]}$  &  -1.52 &  -1.44 &  -1.49 &  -1.39 &  -1.46 &  -1.49 &  -1.52 &  -1.51 &  -1.42 & -1.47$\pm$0.02& 7.50\\
${\rm [Co/Fe]}$ &  -0.15 &  -0.08 &  -0.11 &  -0.30 &  -0.12 &    -   &  -0.05 &    -   &    -   & -0.14$\pm$0.04& 4.93\\
${\rm [Ni/Fe]}$ &  -0.29 &  -0.25 &  -0.26 &  -0.28 &  -0.24 &  -0.34 &  -0.31 &  -0.37 &  -0.27 & -0.29$\pm$0.01& 6.26\\
${\rm [Cu/Fe]}$ &  -0.71 &  -0.88 &  -0.76 &  -0.86 &  -0.80 &    -   &    -   &  -0.72 &  -0.74 & -0.78$\pm$0.03& 4.19\\
${\rm [Zn/Fe]}$ &  -0.04 &  -0.02 &  -0.30 &  -0.13 &  -0.20 &  -0.30 &  -0.30 &  -0.31 &  -0.22 & -0.20$\pm$0.04& 4.61\\
${\rm [Y/Fe]}$  &  -0.79 &  -0.67 &  -0.74 &  -0.65 &  -0.63 &  -0.75 &  -0.80 &  -0.80 &  -0.69 & -0.72$\pm$0.02& 2.25\\
${\rm [Zr/Fe]}$ &  -0.18 &  -0.27 &  -0.15 &  -0.26 &  -0.22 &  -0.16 &  -0.12 &  -0.09 &  -0.09 & -0.19$\pm$0.02& 2.56\\
${\rm [Ba/Fe]}$ &  -0.53 &  -0.42 &  -0.46 &  -0.46 &  -0.41 &  -0.42 &  -0.55 &  -0.45 &  -0.47 & -0.46$\pm$0.02& 2.34\\
${\rm [La/Fe]}$ &  -0.33 &  -0.25 &  -0.26 &  -0.26 &  -0.25 &  -0.23 &  -0.31 &  -0.27 &  -0.31 & -0.27$\pm$0.01& 1.26\\
${\rm [Ce/Fe]}$ &  -0.54 &  -0.60 &  -0.54 &  -0.59 &  -0.64 &  -0.59 &  -0.56 &  -0.50 &  -0.65 & -0.58$\pm$0.02& 1.53\\
${\rm [Pr/Fe]}$ &    -   &  -0.15 &  -0.18 &  -0.23 &  -0.13 &    -   &    -   &    -   &    -   & -0.17$\pm$0.02& 0.71\\
${\rm [Nd/Fe]}$ &  -0.42 &  -0.39 &  -0.40 &  -0.38 &  -0.43 &  -0.43 &  -0.42 &  -0.35 &  -0.37 & -0.40$\pm$0.01& 1.59\\
${\rm [Sm/Fe]}$ &    -   &  -0.11 &    -   &  -0.08 &  -0.04 &  -0.02 &    -   &    -   &  -0.12 & -0.07$\pm$0.02& 0.96\\
${\rm [Eu/Fe]}$ &  -0.23 &  -0.13 &  -0.20 &  -0.18 &  -0.07 &  -0.16 &  -0.28 &  -0.16 &  -0.26 & -0.19$\pm$0.02& 0.52\\
${\rm [Dy/Fe]}$ &    -   &  -0.60 &    -   &  -0.37 &  -0.36 &    -   &  -0.18 &  -0.28 &  -0.19 & -0.33$\pm$0.06& 1.10\\
${\rm [Pb/Fe]}$ &  -0.20 &  -0.21 &  -0.34 &  -0.40 &  -0.25 &  -0.27 &  -0.30 &  -0.12 &  -0.19 & -0.25$\pm$0.03& 1.98\\
\hline
\enddata
\label{t2}
\end{deluxetable}

\section{Results and comparison with literature}

In the following sections, we will discuss in detail our results.
In addition, we will compare them with the literature,
and specifically with \citet{Bu90}, \citet{Sa97}, \citet{DC92}, \citet{Fr97}, \citet{Br97},
\citet{Do11}. The first two give a photometric metallicity, while the
third and the fourth give a spectroscopic metallicity using low-resolution
spectra. \citet{DC92} use the Ca triplet method, while \citet{Fr97}
use a global-fitting of the 4780$\div$5300 \AA\ region including all
the spectral lines. The fifth gives a metallicity based on high-resolution
spectra including [Fe/H] and [O/Fe]. \citet{Bu90} and \citet{Do11}
also discuss the age of the cluster.

\subsection{Iron-peak and $\alpha$ elements}

We found a mean [Fe/H] value for the cluster of:

\begin{center}
{\rm [Fe/H]=-1.47$\pm$0.02 dex}
\end{center}

All the other Fe-peak elements (Sc,V,Cr,Mn,Co,\\ Ni,Cu, and Zn) are underabundant
with respect to Fe, with a range for [El/Fe] between -0.14 dex (Cr and Co) and -0.78 dex (Cu).
The chemical abundances for the $\alpha$ elements O, Mg, Si, Ca, and Ti listed
in Table \ref{t2} are solar scaled or slightly underabundant.
If we use Mg, Si, Ca, and Ti to estimate a mean $\alpha$-element value (O will
be treated separately) we obtain:

\begin{center}
{\rm [$\alpha$/Fe]=-0.02$\pm$0.01 dex}
\end{center}

We conclude that the cluster is solar-scaled as far as $\alpha$-elements are
concerned, in agreement with \citet{Br97}, and at odds with all the other Galactic
GCs of low metallicity. However this behavior is common among
extragalactic objects. We will further discuss $\alpha$ and iron-peak
elements in section 7.

Our results permit us also to explain the disagreement of metallicities
in the literature. Many authors (\citealt{Bu90}, [Fe/H]$\sim$-2; \citealt{Sa97},
[Fe/H]=-1.90; \citealt{DC92}, [Fe/H]=-1.69; \citealt{Fr97}, [Fe/H]=-1.9)
who obtained a metallicity based on photometry (slope and/or color of the RGB) or low-resolution
spectroscopy (Ca triplet or comparison with low-resolution synthetic spectra)
implicitly assumed a typical halo alpha-enhancement, as in all other Galactic GCs.
This fact led them to underestimate the cluster metallicity because
Ruprecht 106 simply is not $\alpha$-enhanced, so, for a given metallicity
([Fe/H]) the RGB-slope is larger or low resolution spectra are apparently
more metal-poor (e.g. the EQW of the Ca triplet is smaller) with respect to
any other GC with similar iron content. \citet{Br97} ([Fe/H]=-1.45) instead obtained a
more reliable result because they measured directly Fe lines. Their value is
in excellent agreement with our.

\subsection{Neutron-capture elements}

\begin{figure}
\epsscale{1.00}
\plotone{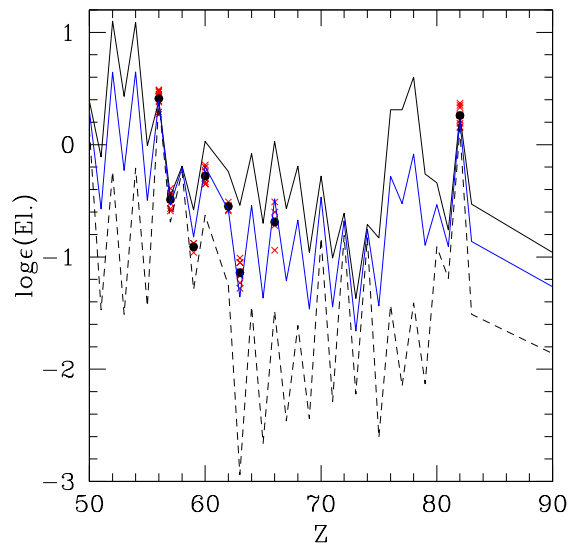}
\caption{Abundance pattern of heavy neutron-capture elements.
Red points represent single stars, while black points the mean abundance for
the cluster. Continuous and dashed lines are pure r and s process patterns
respectively. Continuous blue line is the best fit to our data.}
\label{f3}
\end{figure}

We measured neutron-capture elements from Y to Pb. They 
are produced through the capture of a neutron by a iron-peak seed nucleus.
Once captured, the neutron decays into a proton, and a new nucleus with higher atomic
number is formed.
The capture can is considered slow if the timescale for the neutron capture is large
compared to the timescale of the nuclear decay. In this case we have the
s-process. In the case that the neutron capture is rapid compared with the
timescale of the decay, we have the r-process.
The distribution of neutron-capture elements is different in the two cases,
and a study of their relative abundance gives
information on the relative importance of the two processes on the
contamination of the gas the cluster was formed from.
Because the s-process happens in low-mass AGB stars (1.5-3 M$_{\odot}$,
\citealt{Bu01}), intermediate mass AGB stars (4-8 M$_{\odot}$,
\citealt{Ka07}), and possibly also in massive stars (M$>$20 M$_{\odot}$,
\citealt{Pi08}), while the r-process most
probably occurs in SNeII explosions, abundances of neutron-capture
elements tell us how these kind of stars contributed to the formation of the
cluster.
For this purpose in Fig.~\ref{f3} we plotted the abundance of Ba, La, Pr, Nd,
Sm, Eu, Dy, and Pb. On our data we superposed the two abundance curves 
of the pure s (dashed black line) and r (continuous black line) process taken from \citet{Sn08}. Ba and
Pb were used to set the
zero-point of the pure s and pure r process curves (see below), while La, Pr, Nd, Sm, Eu,
and Dy are those elements more sensitive to the s and r processes. 
The Ce abundance does not
follow any curve in Fig.~\ref{f3}, probably because
of some non-negligible systematic error in our abundance estimation for this
element, so we decided not to use it in our investigation.
The two curves were shifted in the y direction in order to match the Ba and Pb
abundances that are the same for the two processes. Abundances of the other
elements are in between the two curves and the best fitting line (the blue
continuous line) indicates  that r-process contributed 66\% and s-process
34\% to the cluster abundances. This result is confirmed by the mean
log$\epsilon$(La/Eu) of the cluster, that is:

\begin{center}
{\rm log$\epsilon$(La/Eu)=0.65$\pm$0.02}
\end{center}

According to \citet[Fig. 4]{Ro09}, this corresponds to a contribution of 71\% 
from the r-process. We can summarize these results by estimating that r-process contributed
a 68.5$\pm$2.5\% and s-process a 31.5$\pm$2.5\% to the contamination
of the primordial gas the cluster was formed from.

\begin{deluxetable}{lccccccc}
\tablecolumns{8}
\tablewidth{0pc}
\tablecaption{Estimated errors on abundances due to errors on atmospheric
parameters and to spectral noise compared with the observed errors.}
\tablehead{
ID & \colhead{$\Delta$T$_{\rm eff}$=30 K}  & \colhead{$\Delta$log(g)=0.09} & 
\colhead{$\Delta$v$_{\rm t}$=0.03 km/s}  & \colhead{$\Delta$[Fe/H]=0.05} & S/N
& \colhead{$\sigma_{\rm tot}$} & \colhead{$\sigma_{\rm obs}$}
}
\startdata
\hline
$\Delta$([O/Fe])  & 0.03 & 0.06 & 0.03 & 0.02 & 0.02 & 0.08 & 0.06$\pm$0.02\\
$\Delta$([Na/Fe]) & 0.01 & 0.01 & 0.00 & 0.01 & 0.05 & 0.05 & 0.05$\pm$0.01\\
$\Delta$([Mg/Fe]) & 0.02 & 0.01 & 0.00 & 0.01 & 0.05 & 0.06 & 0.04$\pm$0.01\\
$\Delta$([Al/Fe]) & 0.01 & 0.00 & 0.01 & 0.01 & 0.05 & 0.08 & -            \\
$\Delta$([Si/Fe]) & 0.03 & 0.01 & 0.01 & 0.01 & 0.05 & 0.06 & 0.05$\pm$0.01\\
$\Delta$([Ca/Fe]) & 0.00 & 0.00 & 0.00 & 0.00 & 0.04 & 0.04 & 0.05$\pm$0.01\\
$\Delta$([Sc/Fe]) & 0.04 & 0.05 & 0.02 & 0.01 & 0.01 & 0.07 & 0.07$\pm$0.02\\
$\Delta$([Ti/Fe]) & 0.02 & 0.00 & 0.00 & 0.01 & 0.03 & 0.04 & 0.04$\pm$0.01\\
$\Delta$([V/Fe])  & 0.02 & 0.01 & 0.00 & 0.01 & 0.08 & 0.08 & 0.06$\pm$0.02\\
$\Delta$([Cr/Fe]) & 0.02 & 0.00 & 0.00 & 0.00 & 0.03 & 0.04 & 0.03$\pm$0.01\\
$\Delta$([Mn/Fe]) & 0.01 & 0.01 & 0.01 & 0.00 & 0.04 & 0.04 & 0.05$\pm$0.01\\
$\Delta$([Fe/H])  & 0.04 & 0.01 & 0.01 & 0.01 & 0.01 & 0.05 & 0.05$\pm$0.01\\
$\Delta$([Co/Fe]) & 0.01 & 0.00 & 0.00 & 0.01 & 0.08 & 0.08 & 0.09$\pm$0.03\\
$\Delta$([Ni/Fe]) & 0.00 & 0.00 & 0.00 & 0.00 & 0.02 & 0.02 & 0.04$\pm$0.01\\
$\Delta$([Cu/Fe]) & 0.01 & 0.00 & 0.00 & 0.01 & 0.05 & 0.05 & 0.07$\pm$0.02\\
$\Delta$([Zn/Fe]) & 0.06 & 0.02 & 0.01 & 0.01 & 0.08 & 0.10 & 0.11$\pm$0.03\\
$\Delta$([Y/Fe])  & 0.04 & 0.04 & 0.02 & 0.00 & 0.08 & 0.10 & 0.07$\pm$0.02\\
$\Delta$([Zr/Fe]) & 0.05 & 0.04 & 0.02 & 0.02 & 0.04 & 0.08 & 0.07$\pm$0.02\\
$\Delta$([Ba/Fe]) & 0.02 & 0.05 & 0.02 & 0.03 & 0.01 & 0.07 & 0.05$\pm$0.01\\
$\Delta$([La/Fe]) & 0.04 & 0.03 & 0.02 & 0.00 & 0.02 & 0.06 & 0.04$\pm$0.01\\
$\Delta$([Ce/Fe]) & 0.05 & 0.04 & 0.01 & 0.01 & 0.04 & 0.08 & 0.05$\pm$0.01\\
$\Delta$([Pr/Fe]) & 0.02 & 0.04 & 0.01 & 0.00 & 0.05 & 0.07 & 0.04$\pm$0.02\\
$\Delta$([Nd/Fe]) & 0.02 & 0.03 & 0.01 & 0.00 & 0.02 & 0.04 & 0.03$\pm$0.01\\
$\Delta$([Sm/Fe]) & 0.02 & 0.05 & 0.03 & 0.02 & 0.07 & 0.10 & 0.04$\pm$0.01\\
$\Delta$([Eu/Fe]) & 0.04 & 0.05 & 0.03 & 0.02 & 0.02 & 0.08 & 0.07$\pm$0.02\\
$\Delta$([Dy/Fe]) & 0.01 & 0.07 & 0.02 & 0.01 & 0.02 & 0.08 & 0.15$\pm$0.04\\
$\Delta$([Pb/Fe]) & 0.04 & 0.01 & 0.01 & 0.02 & 0.03 & 0.06 & 0.09$\pm$0.02\\
\hline
\enddata
\label{t3}
\end{deluxetable}

\section{Ruprecht 106 as the first single population GC}

\begin{figure*}[ht!]
\epsscale{2.00}
\plotone{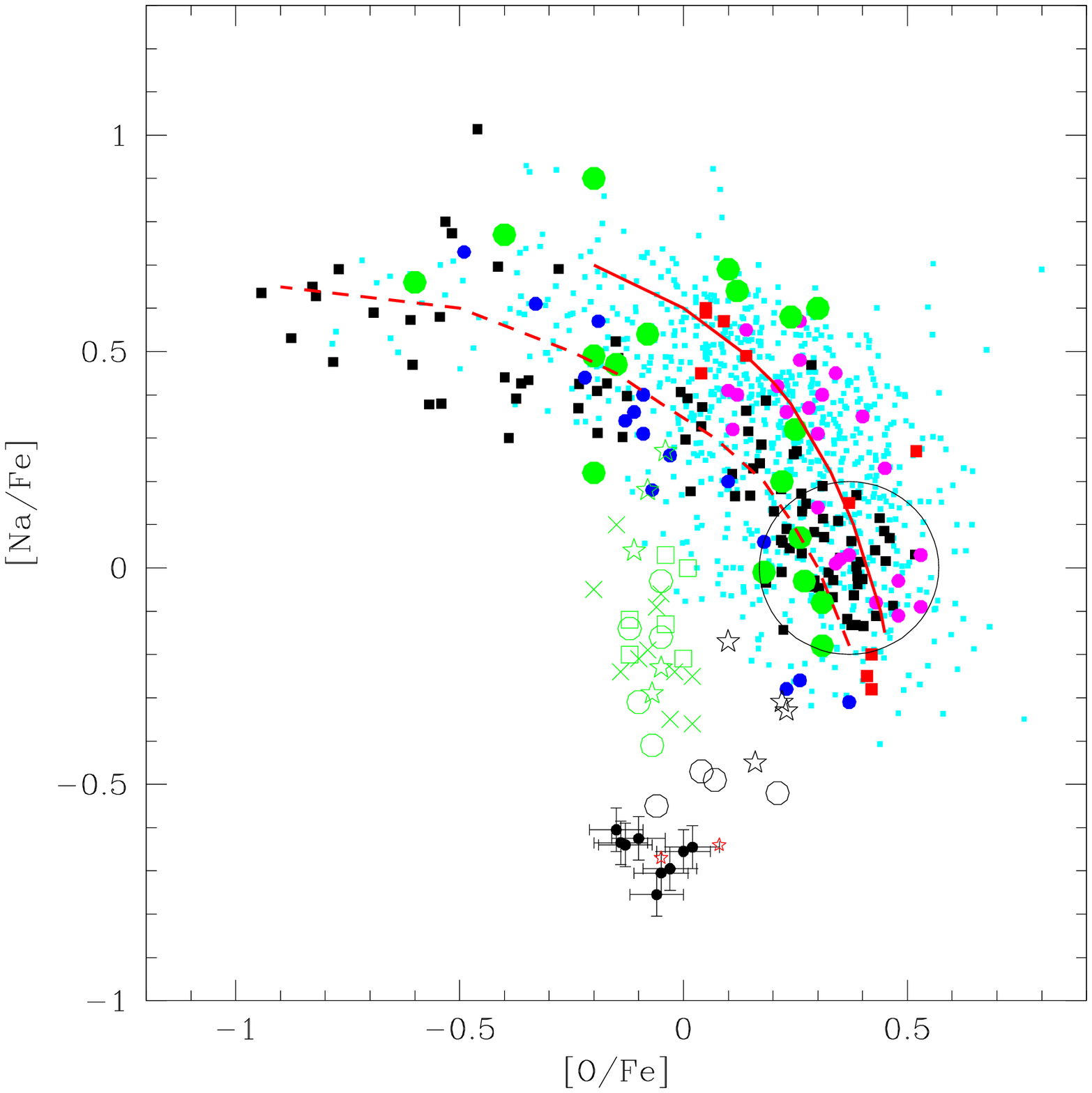}
\caption{[Na/Fe] vs. [O/Fe] in GCs. Filled cyan squares: GCs from
\citet{Ca09}. Filled blue circles: NGC~1851. Filled black squares:
NGC~2808. Filled red squares: M22. Filled magenta circles: M4. Filled green
circles: Old LMC clusters. Open green symbols: intermediate-age LMC clusters. 
Open black stars: Terzan 7. Filled black circles: Ruprecht 106 (this paper). Open red
stars: Ruprecht 106 by \citet{Br97}. See text for more details.}
\label{f4}
\end{figure*}

As mentioned in the Introduction, other
GCs have been proposed as single population objects, like Palomar 3 \citep{Ca11},
due to the small extension of its red HB. \citet{Ko09} studied
spectroscopically four bright RGB stars in this cluster, but due to the low
S/N and the small statistics they could not prove the presence or absence of
a spread in light elements, and so they could not confirm or
reject the \citet{Ca11} hypothesis.
The main aim of this paper is to verify if Ruprecht 106 might be the first
bonafide example of a single population old Galactic GC.
For this purpose we report in Fig.~\ref{f4} the Na-O abundances of our 9 stars
(black points with errorbars). In this figure, for comparison, we report also
data for the GCs studied in \citet[filled cyan squares]{Ca09}, NGC~1851 \citep[filled blue
circles]{Vi10}, NGC~2808 \citep[filled black squares]{Ca06}, M4 \citep[open magenta
circles]{Vi11}, M22 \citep[open red squares]{Ma09}, old (filled green
circles) and intermediate-age (other open green symbols) clusters in the LMC studied by
\citet{Mu08,Mu09}, and the Sagittarius cluster Terzan 7 \citep[open black stars]{Sb07}.
Open red stars represent the two targets in Ruprecht 106 studied by \citet{Br97}. These stars are
\#Ru$_{\rm 1614}$ and \#Ru$_{\rm 2004}$ and the authors find
[O/Fe]=-0.05,+0.08 and [Na/Fe]=-0.47,-0.44 respectively. Their Na abundance is based
on the 8190 \AA\ doublet so, according to \citet{Ma00}, a NLTE correction of
$\sim$-0.2 dex looks appropriate. The agreement with our results is good.

Ruprecht 106 stars define a clump at [O/Fe]=-0.07 dex and [Na/Fe]=-0.66 dex
that has no intrinsic dispersion. This is confirmed by
Tab.~\ref{t3}, where the theoretical spread ($\sigma_{\rm tot}$)  matches very
well within the errors with the observed spread ($\sigma_{\rm obs}$) both for
O and Na. This is true also for the other elements.
This confirms our initial hypothesis that  Ruprecht 106 is the first
example of a single population GC, although it is possible that nature has
conspired against us and we simply did not detect a true spread. See below
for further discussion of this point. Recall that the Na-O
anticorrelation has been used to define a GC \citep{Ca09}, so its
absence in Ruprecht 106 is of particular importance.

However Fig.~\ref{f4} has several other interesting implications. 
First of all Ruprecht 106 occupies a totally unique position in this diagram
(see Section 7 for further discussion).
In addition, among Galactic GCs at least two distinct Na-O anticorrelations appear, that are
plotted as red lines. One is O-richer (continuous line) and represented by the
trend of the stars in M4 and M22, while the other is O-poorer (dashed line) and
represented by the trend of the stars in NGC~1851 and NGC2808. This fact can be
explained by the $\alpha$-enhancement a cluster was formed with and remembering
that O is an $\alpha$ element too. 
If first generation stars in a cluster were born relatively O-poor (with
respect to the continuous line of Fig.~\ref{f4}), second generation stars
will also share the same chemical behavior.

As an example, NGC~1851 and M4 have a difference
in their $\alpha$-enhancement of 0.07 dex (as defined by the mean abundance of
Si and Ca), while the difference between their O content is $\sim$0.1 dex
among the entire Na-O anticorrelation. So the O content of second generation
stars does not depend only on the internal chemical evolution of the clusters,
but also on the primordial O abundance of the progenitor cloud.

Old LMC clusters also follow the general Galactic trend, but this is not the case for
intermediate-age LMC clusters and Terzan 7 which do not show an appreciable spread in O. On the other
hand three of them show a possible spread in Na (NGC~1651: open green circles,
NGC~1978: green crosses, NGC~2173: open green stars). NGC~1978 also shows a
hint of a Na-O anticorrelation. Only two clusters (NGC~1783: open green squares) and Terzan 7 have a
small dispersion in Na, possibly compatible with zero, but still larger then
Ruprecht 106, but all of these samples are relatively small.

In Fig.~\ref{f4} the region presumably inhabited by first generation stars in Galactic
and LMC globular cluster is indicated by a large black circle.
Even if Ruprecht 106 does not fit at all with any known Milky Way or LMC cluster,
we can still state that these stars still correspond to the first generation.
This is because this first generation is characterized by the fact that
[O/Fe]=[$\alpha$/Fe] within 0.1 dex (where $\alpha$ is defined as 
the mean abundance of Si and Ca). Second generations stars
have [O/Fe]$<$[$\alpha$/Fe]. As an example, in M4 stars of the first generation ([$\alpha$/Fe]=+0.42)
have [O/Fe]=+0.42 dex, while stars of the second have [O/Fe]=+0.25 (the
difference with respect the $\alpha$-enhancement is -0.17 dex).
This behavior is expected because the first generation was
born from fresh material, enhanced in all $\alpha$-elements,
including O. Instead the second generation was born from O-depleted material,
which however had other $\alpha$-elements untouched.
For Ruprecht 106 we found that the O content is only slightly lower than other
$\alpha$-elements (0.07 dex), so we can say that the observed stars correspond to the first
generation.
With this information we can answer the following question: what is the
probability that we have missed a second generation if present? \citet{Ca09}
found that $\sim$30-40\% of stars now remaining in a GC belong to the first generation. We observed
9 stars and all belong to it. So the probability that we have
missed the second population is:

\begin{center}
{\rm P$\sim$0.35$^{9}$=0.00008=0.008\%}
\end{center}

This probability is low enough to confidently state that we did not miss any
cluster sub-population. This would be true (however not in as clear a way)
also if our stars had belonged to the second generation, for which we would
obtain P=2\%. However, we recognize that our sample is still small,
especially compared to the samples generally used to define the Na-O
anticorrelation often with $>$100 stars. Spectra of additional stars in Ruprecht
106 would be most welcome.

Another relevant piece of evidence comes from the HB.
It has a dispersion in color of $\sim$0.2 mag (see Fig.~\ref{f1}).
According to \citet{Ca11} a dispersion of 0.3 mag or lower is an indication
that no multiple populations are present in a cluster. So this fact further
reinforces our main result.

The last possibility could be that Rup 106 represents one of those
first generation-mainly clusters, where a second generation is present but not dominant
\citep{Ca11}. However in this case we should still see a HB extended to the
blue because second generation HB stars populate a hotter and bluer HB part
than first generation stars as discussed in the introduction. In Rup 106 this
would correspond to HB stars with (V-I)$<$0.5 (see Fig.~\ref{f1}), where no
stars at all are present. 
On the other hand the chances that Rup 106 is such an object are decreased by
the fact that first generation-mainly clusters are assumed to not have filled
their tidal radius and thus did not lose many first generation stars. However,
odds with this, Rup 106 is known to be a tidally-filling cluster \citep{Ba11}.
In conclusion, the evidence strongly points to Rup 106 asa first generation
only cluster.

Considering the above factors, we conclude that it is very likely that
Ruprecht 106 is the first confirmed example of a single population, globular cluster.

What about other possible candidates?
One is Palomar 3. \citet{Ca11} suggested it as a single
population cluster based on its HB, that has a color spread of 0.25 mag (see
their Fig. 2). For Ruprecht 106 instead we found a value $\leq$0.20 mag (see Fig.~\ref{f1}).
On the spectroscopic side, \citet{Ko09} found that the observed Na spread in
Palomar 3 slightly exceeds that expected from theoretical errors and is not accompanied
by a spread in O. This is not surprising because Palomar 3 stars are
located in a region of the Na-O anticorrelation where little or no O variation
is expected.
In our opinion, these results together suggest that
Palomar 3 has a small but real spread in light elements, and it is therefore
probably not an example of a single population GC. In addition only 4 stars
have been observed at high resolution. 
On the other hand, NGC~1783 and Terzan 7 have a dispersion in Na of 0.10
and 0.12 dex respectively, larger then Ruprecht 106, but that could be due to
larger internal errors. So they could be single population GCs.
In any case for these three clusters further investigation is required to
reveal their nature. Obviously, it is of interest to observe other possible
candidate single-population GCs.

We finally note that these candidates are extragalactic or, in the case of
Palomar 3, probably have an extragalactic origin \citep{Ca11}. This is true
also for Ruprecht 106 (see next section). A possible reason could be that all
low mass single population Galactic GC were destroyed due to the tidal
interaction with the Milky Way, or simply they did not form at all.
However, most low mass GCs have not been adequately investigated yet.

\begin{figure}[ht!]
\epsscale{1.10}
\plotone{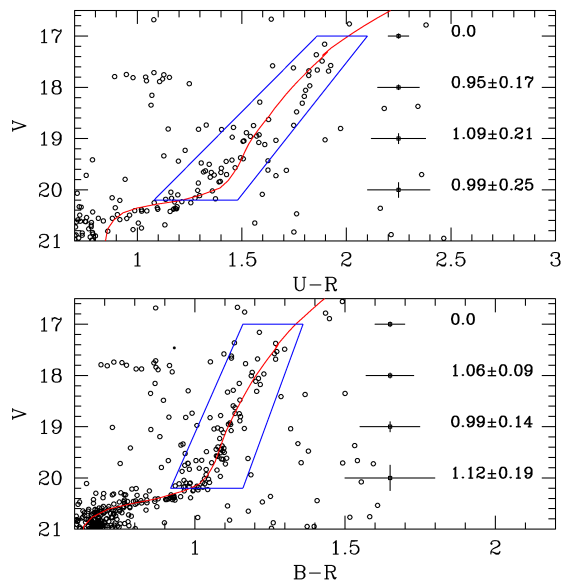}
\caption{Upper pannel: V vs. U-R CMD of the cluster zoomed on the RGB.
Lower pannel: V vs. B-R CMD of the cluster zoomed on the RGB.}
\label{f5b}
\end{figure}

\begin{figure}[ht!]
\epsscale{1.00}
\plotone{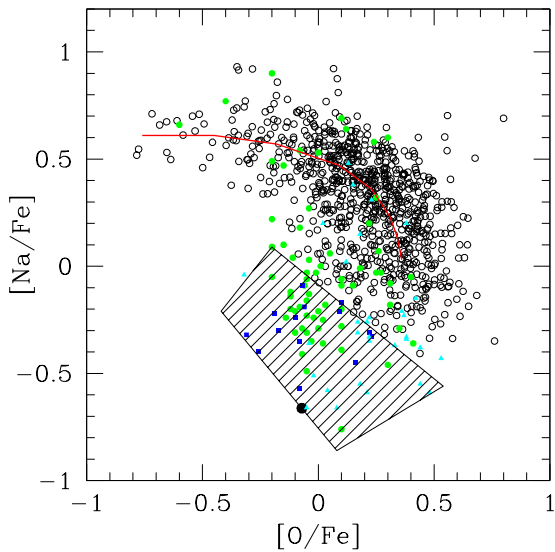}
\caption{Na-O anticorrelation. Filled cyan triangles: 
Draco, Sextans, and Ursa Minor dwarf galaxies. Filled blue squares: 
Sagittarius dwarf galaxy. Filled green circles: LMC. Open black
circles: GCs from \citet{Ca09}. Filled black circle: Ruprecht 106.
See text for more details.}
\label{f5}
\end{figure}

\section{Additional photometric evidence}

It now well known that the U filter is able to disentagle multiple populations along
the RGB of Globular Clusters \citep{Ma08,Sb11} due to its sensitivity to
chemical inhomogeneities. 
For this purpose we extracted photometric material for 
Ruprecht 106 from the ESO public archive
\footnote{http://archive.eso.org/eso/eso\_ archive\_ main.html}.
It consists of a series of images in UBVRI taken with the SUperb Seeing Imager (SUSI2) at
La Silla Observatory on the night of July 21, 2002.
The detector has a scale of 0.08 arcsec/pixel, allowing to cover
5.5$\times$5.4 arcmin on the sky.

This data-set has the widest optical wavelength coverage for Ruprecht 106 available, and it is
ideal to study the detailed shape of the Red Giant Branch. The night was
photometric according to the weather report from ESO La Silla, with an average
seeing of 0.8 arcsec in the V pass-bands. Exposures range from 30 to 900 in U,
B, V, and R. The cluster was also observed in I, but the images show 
significant fringing, which made it impossible to extract good
photometry. Only the north-east portion of the cluster was surveyed.
After pre-reduction (bias and flat-field corrections) images have been reduced
using the DAOPHOT/ALLSTAR routines. SUSI2 has two detectors, which have been
separated and reduced individually. The two extracted catalogs have then
been merged. 
To tie the instrumental photometry to the standard system we used  the average
zero-points and color terms reported in the website of the instrument.
\footnote{http://www.eso.org/sci/facilities/lasilla/}.

In Fig.~\ref{f5b} we present a zoom on the RGB region in 2 different 
color combinations: V vs. U-R in the upper panel, and V vs. B-R in the lower, 
to maximize the color range of the stars. The V vs. B-R combination was
included as a reference only, because colors other than U are not as sensitive to
chemical inhomogeneities.
We use the metallicity derived in this paper to fit the star distribution
with theoretical isochrones (red lines) from the Padova database \citep{Mar08}.
We find that the best-fitting parameters are an age of 12 Gyrs, a reddening E(B-V) =0.19,
and an apparent distance modulus (m-M)$_{\rm V}$= 17.20 mag.  Both reddening and
distance modulus are in good agreement with the literature value.

Color-coded in blue is the region where bona-fine 
RGB stars are located. Lacking any quantitative membership, these have been 
selected as the RGB stars lying within the cluster core region (core radius=1.0 arcmin, \citealt{Ha96}). 
Photometric errors, in magnitude and color, are also indicated with 
error-bars color-coded in black. 
To investigate a possible intrinsic photometric spread in the RGB, we compared
its observed color scatter with the natural width expected from considering the
photometric errors only. These are calculated as 

\begin{center}
$\sigma_{B-R}=2\times \sqrt(\sigma_B^2 + \sigma_R^2)$
$\sigma_{U-R}=2\times \sqrt(\sigma_U^2 + \sigma_R^2)$
\end{center}

depending on the CMD. As for the observed color scatter at a given V, we simply
consider the difference in color  $\Delta (B-R)$, and $\Delta (U-R)$ between
the two most separated stars in the RGB at about the same V. This is because 
the RGB is not very rich, and therefore any other statistical calculation
would be not robust. 
The results for 4 different V magnitudes are reported directly 
in the figure as $\sigma$/$\Delta$ with their errors. 
Apart from the value at V = 17, where no RGB stars are located, the 
ratios are compatible with 1 within the errors, indicating that the RGB is not
wider than the amount expected from photometric errors only. 
We therefore conclude that this photometric dataset supports 
the spectroscopic result that no significant spread exists among 
Ruprecht 106 RGB stars.

We can further compare our (U-R) vs. V CMD with that for M4 \citep[(U-B) vs
U, Fig. 11]{Ma08}. M4 has a very broad RGB with a spread of $\sim$0.2
mag. in color. At odds with this, the Rup 106 RGB spread is only 0.1
mag. Because M4 has one of the lowest Na-O spreads among GCs
\citep{Ca09}, we conclude that this photometry further supports our
spectroscopic result.

\section{Comparison with Galactic and extragalactic environments}

The position of Ruprecht 106 stars in Fig.~\ref{f4} is extraordinary. They do not fit
any Na-O trend defined by Galactic GC.
This is an indication that Ruprecht 106 has an extragalactic origin.
In order to investigate this point more deeply, we compare our results with
Galactic and extragalactic environments.
For this purpose, we plotted in Fig.~\ref{f5} data for Galactic GCs
\citep[open black circles]{Ca09} to trace the Na-O trend for the Galactic
Halo. In addition we added data for the LMC field and cluster stars \citep[green filled
circles]{Jo06,Po08,Mu08,Mu09}, Sagittarius dwarf galaxy field and cluster
stars \citep[filled blue squares]{Sb07}, the Draco, Sextans, and Ursa
Minor dwarf galaxy field stars \cite[filled cyan triangles]{Sh01}.
The continuous red line indicates the mean trend for a Galactic GCs. The main result
is that, while the area occupied by the Galactic stars is common also to
the extragalactic objects, there is a region occupied only by extragalactic
stars, indicated by the shaded region, that are more Na-poor with respect to the
Galactic Halo.
Ruprecht 106 (the filled black circle) lies at the opposite extreme of this
area, and because of this it can be considered with high confidence an extragalactic
object.

\begin{figure}[ht!]
\epsscale{1.00}
\plotone{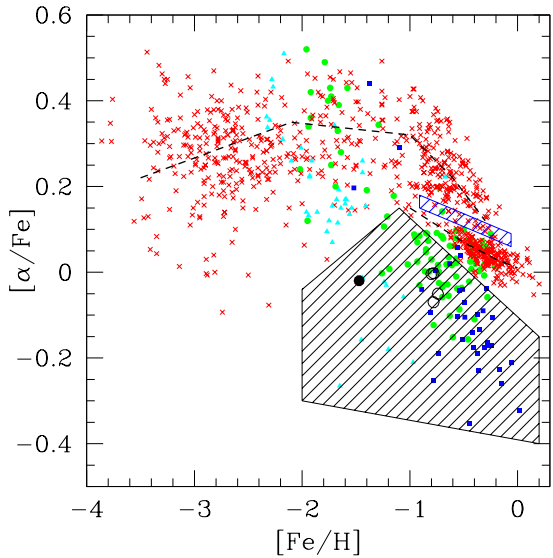}
\caption{[$\alpha$/Fe] vs. [Fe/H]. Filled cyan triangles: 
Draco, Sextans, and Ursa Minor dwarf galaxies. Filled blue squares: 
Sagittarius dwarf galaxy. Filled green circles: LMC. Red
crosses: Milky Way. Filled black circle: Ruprecht 106.
See text for more details.}
\label{f6}
\end{figure}

Additional support for an extragalactic origin comes from Fig.~\ref{f6}. Here we plot the [$\alpha$/Fe]
vs. [Fe/H] relation for extragalactic stars defined as in Fig.~\ref{f5} and
galactic field stars (red crosses) from the following sources: 
\citet{Fu00,Ca04,Re03,Ba05,Re06}. In this plot we added also LMC stars from
\citet{Mo05}.
[$\alpha$/Fe] was defined as the mean abundance of Mg, Si, Ca, and Ti except
for \citet{Po08,Ba05,Mo05} for which we used Mg, Ca, and Ti. For \citet{Mu09}
we were forced to use the only $\alpha$-element available (besides O), i.e. Mg.
In any case, stars from \citet{Mu09} follow the general trend of the LMC.
First of all, in Fig.~\ref{f6} we identified a double trend of [$\alpha$/Fe]
vs. [Fe/H] for the Galaxy, where stars define a first continuous path from
[Fe/H]$\sim$-3.5 dex up to  [Fe/H]$\sim$-0.2 dex, indicated by the upper black
dashed line. The second continuous path goes from [Fe/H]$\sim$-1.2 up to
[Fe/H]$\sim$+0.2 and is indicated by the lower black dashed line.
Between the two trends there is an almost empty region indicated by the blue
shaded area. This result is not new and was recently noticed by \citet{Ad11}.
These authors found two distinct [$\alpha$/Fe] vs. [Fe/H] trends among
Galactic stars, with an almost empty region between them (see their Fig. 1),
exactly as in our case.
Apart from that, we can see that extragalactic stars have the same mean
$\alpha$-enhancement as the Milky Way up to [Fe/H]$\sim$-1.5. More metal rich
extragalactic stars instead tend to have a much lower $\alpha$ content than
Galactic stars of the same metallicity.
The black shaded area indicates the region populated only by
extragalactic objects.
Again Ruprecht 106 (the filled black circle) lies in this area well removed
from any Galactic star. 

\begin{figure}[ht!]
\epsscale{1.00}
\plotone{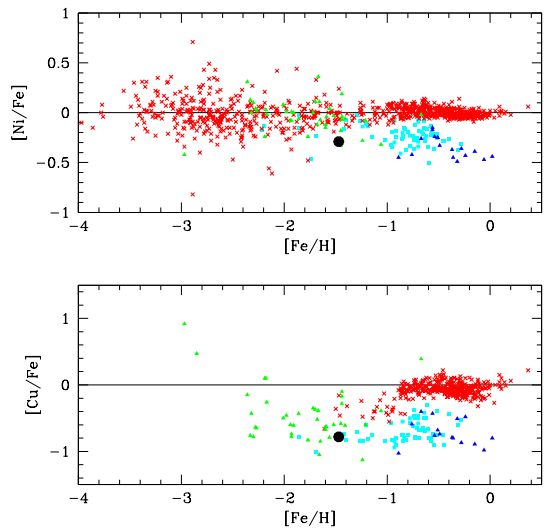}
\caption{[Ni/Fe] vs. [Fe/H] and [Cu/Fe] vs. [Fe/H]. Filled cyan triangles: 
Draco, Sextans, and Ursa Minor dwarf galaxies. Filled blue squares: 
Sagittarius dwarf galaxy. Filled green circles: LMC. Red
crosses: Milky Way. Filled black circle: Ruprecht 106.
See text for more details.}
\label{f7}
\end{figure}

Finally in Fig.~\ref{f7} we compare the Ni and Cu content of Rup 106 with
Galactic and extragalactic stars. Literature sources are those discussed above.
Solar-scaled abundances are shown as continuous black lines for reference.
As far as Ni is concerned, Galactic stars follow a solar-scaled trend with a
large spread for [Fe/H]$<$-2. On the other hand, extragalactic objects are
solar scaled below [Fe/H]=-1.5. Then they start to deviate, reaching
[Ni/Fe]$\sim$-0.5 for solar metallicity. Rup 106 is located at the metallicity
where the deviation starts, but below the Galactic trend and fully compatible
with extragalactic targets.

As for Cu, Galactic stars follow a solar scaled trend down to
[Fe/H]$\sim$-0.9. Below that metallicity they drop to
[Cu/Fe]$\sim$-0.4. Extragalactic targets are more Cu poor on average with
$<$[Cu/Fe]$>$$\sim$-0.6 regardless of the iron content. Although the number of
Galactic stars at the metallicity of Ruprecht 106 is very small, our 
cluster is again located below the Galactic trend and fully compatible with
extragalactic targets, once again pointing to an extragalactic origin.

It is of course very difficult to say where Ruprecht 106
comes from. It was suggested initially to be part of the  Sagittarius dwarf
galaxy \citep{Be03}, but \citet{La10} discarded this hypothesis because it does not belong
to any possible stream of this system, so we regard its exact origin as an
open question, but an extragalactic formation is strongly favoured by our data.

\section{Summary and conclusions}

In this paper we present detailed chemical abundances of red giants in the globular
cluster Ruprecht 106. We studied 29 elements from C to Pb, including light,
$\alpha$, iron-peak, and neutron-capture. Our main aim was to investigate if
Ruprecht 106 is a single population GC.
All GCs studied up to now show some spread in their chemical abundances. There are a
few examples where no spread was found, but those results are doubtful due to the
small sample of stars observed and to the lack of an accurate error analysis. 
We analyzed 9 member stars and performed an accurate error abundance analysis.
We found that the observed spread in all elements, in particular Na and O, is
totally within the measurement errors.
We calculated also a negligible probability of having missed stars with an intrinsic
difference in their chemical content with respect to our targets given the
nominal ratio of first to second generation stars.
This is also confirmed by the small color spread of the HB.
No intrinsic abundance spread is present in this GC in any element. Although our sample
is still relatively small and more observations would be of great
interest, our evidence strongly suggests that Ruprecht 106 is the first
genuine old, massive GC with only a single population.

In addition, we could establish the following:

\begin{itemize}
 \item Ruprecht 106 has [Fe/H]$\sim$-1.5 and [$\alpha$/Fe]$\sim$0.0.
 This solves the disagreement in the literature between metallicity measurements based on
 photometry or low resolution spectroscopy and those based on high resolution
 spectroscopy.
 \item Its neutron-capture element abundances point toward a contamination
 of the gas the cluster was formed from by both the s and r processes. The
 contamination fraction is $\sim$30\% and 70\% respectively.
 \item NGC~1783, Terzan 7, and Palomar 3 are candidate single population
 GCs. Available data are uncertain and new more accurate studies are
 required to confirm their nature.
 \item Na/O, $\alpha$, and iron-peak abundances clearly point toward an
 extragalactic origin of the cluster. No progenitor galaxy or stream have been
 clearly identified yet.

\end{itemize}

Ruprecht 106 has a present day mass of M=10$^{4.83}$ M$_{\odot}$. This is not its
initial mass because it lost a fraction of its stars due to internal precesses
and to interaction with the Milky Way. A determination of the initial mass
would require the knowledge of its orbit and a detailed dynamical
simulation. This is beyond the scope of this paper. In any case we can fix
M=10$^{4.83}$ M$_{\odot}$ as a lower limit for the initial mass threshold
below which no ejecta are retained by the gravitational potential of a cluster
and no second generation of stars is formed. Note that NGC~6838, one of the
\citet{Ca09} sample clusters, has a present day mass of M=10$^{4.30}$
M$_{\odot}$, much less than that of Ruprecht 106, but shows a real Na-O
spread. We hypothesize that NGC~6838 was originally more massive than Ruprecht
106, but subsequently lost more mass.

\acknowledgments
S.V. and D.G.gratefully acknowledge support from the Chilean
Centro de Excelencia en Astrof\'\i sica
y Tecnolog\'\i as Afines (CATA) grant PFB-06/2007.
The authors gratefully acknowledge also Ata Sarajedini who kindly provided
the ACS@HST photometry.
S.V. gratefully acknowledges the support provided by FONDECYT N. 1130721


\begin{thebibliography}{}

 \bibitem[Adibekyan et al. (2011)]{Ad11} Adibekyan, V.Z., Santos, N.C., Sousa,
 S.G., \& Israelian, G. 2011, arXiv1111.4936

 \bibitem[Alonso et al. (1999)]{Al99} Alonso, A., Arribas, S. \&
  Mart\'inez-Roger, C. 1999, A\&AS, 140, 261

 \bibitem[Barklem et al. (2005)]{Ba05} Barklem, P.S., Christlieb, N., Beers, T.C., 
  Hill, V., Bessell, M.S., Holmberg, J., Marsteller, B., 
  Rossi, S., Zickgraf, F.J., \& Reimers, D. 2005, A\&A, 439, 129

 \bibitem[Baumgardt et al. (2011)]{Ba11} Baumgardt, H., Lockmann, U. \&
 Kroupa, P.  2011, ASPC, 439, 96

 \bibitem[Bellazzini et al. (2003)]{Be03} Bellazzini, M., Ferraro, F.R., \&
 Ibata, R. 2003, AJ, 125, 188

 \bibitem[Brown et al. (1997)]{Br97} Brown, J.A., Wallerstein, G., \& Zucker, D.
 1997, AJ, 114, 180

 \bibitem[Buonanno et al. (1990)]{Bu90} Buonanno, R., Buscema, G., 
  Fusi Pecci, F., Richer, H.B., \& Fahlman, G.G. 1990, AJ, 100, 1811

 \bibitem[Busso et al. (2001)]{Bu01} Busso, M., Gallino, R., Lambert,
 D.L., Travaglio, C., \& Smith, V.V. 2001, ApJ, 557, 802

 \bibitem[Cayrel et al. (2004)]{Ca04} Cayrel, R., Depagne, E., Spite, M., Hill, V., 
 Spite, F., Fracois, P., Plez, B., Beers, T., Primas, F., Andersen, J.
 2004, A\&A, 416, 1117

 \bibitem[Caloi \& D'Antona (2011)]{Ca11} Caloi, V., \& D'Antona, F.
 2011, MNRAS, 417, 228

 \bibitem[Carraro et al. (2012)]{Ca12} Carraro, G., Villanova, S., Demarque, P., McSwain, M.V., Piotto,
 G., \& Bedin, L.R. 2006, ApJ, 643, 1151

 \bibitem[Carretta et al. (2006)]{Ca06} Carretta, E., Bragaglia, A., 
 Gratton, R.G., Leone, F., Recio-Blanco, A., \& Lucatello, S. 2006, A\&A, 450,
 523

 \bibitem[Carretta et al .(2007b)]{Ca07} Carretta, E., Bragaglia, A., 
 Gratton, R.G., Lucatello, S., \& Momany, Y. 2007, A\&A, 464, 927

 \bibitem[Carretta et al. (2009)]{Ca09} Carretta, E., Bragaglia, A., 
 Gratton, R.G., Lucatello, S., Catanzaro, G., Leone, F., Bellazzini, M., 
 Claudi, R., D'Orazi, V., \& Momany, Y. 2009, A\&A, 505, 117

 \bibitem[Carretta et al. (2010)]{Ca10} Carretta, E., Bragaglia, A., 
 Gratton, R.G., Lucatello, S., Bellazzini, M., Catanzaro, G., Leone, F., 
 Momany, Y., Piotto, G. \& D'Orazi, V. 2010, A\&A, 520,  95

 \bibitem[D'Antona et al. (2002)]{Da02} D'Antona, F., Caloi, V., Montalb\'an, 
  J., Ventura, P., \& Gratton, R. 2002, A\&A, 395, 69

\bibitem[D'Antona 
\& Caloi(2008)]{Da08} D'Antona, F., \& Caloi, V.\ 2008, \mnras, 390, 693 

 \bibitem[de Silva et al. (2009)]{De09} de Silva, G.M.,  Gibson, B.K.,  Lattanzio, J., \& Asplund,
 M. 2009, A\&A, 500, 25

 \bibitem[Decressin et al. (2007)]{De07} Decressin, T., Meynet, G., 
 Charbonnel, C., Prantzos, N., \& Ekstrom, S. 2007, A\&A, 464, 1029

 \bibitem[D'Ercole et al. (2008)]{De08} D'Ercole, A., Vesperini, E.,
  D'Antona, F., McMillan, S.L.W., \& Recchi, S. 2008, MNRAS, 391, 825	

 \bibitem[Da Costa et al. (1992)]{DC92} Da Costa, G.S., Armandroff, T.E.,
  \& Norris, J.E. 1992, AJ, 104, 154


 \bibitem[Dotter et al. (2011)]{Do11} Dotter, A., Sarajedini, A., \& Anderson,
 J. 2011, ApJ, 738, 74

 \bibitem[Francois et al. (1997)]{Fr97} Francois, P., Danziger, J., Buonanno,
 R., \& Perrin, M. N. 1997, A\&A, 327, 121

 \bibitem[Fulbright (2000)]{Fu00} Fulbright, J.P. 2000, AJ, 120, 1841

 \bibitem[Gratton et 
 al.(2004)]{Gr04} Gratton, R., Sneden, C., \& Carretta, E.\ 2004, \araa, 42, 385 

 \bibitem[Gratton et 
 al.(2011)]{Gr11} Gratton, R.~G., Lucatello, S., Carretta, E., et al.\ 2011, \aap, 534, A123 

 \bibitem[Gratton et 
 al.(2012)]{Gr12} Gratton, R.~G., Lucatello, S., Carretta, E., et al.\ 2012, \aap, 539, A19 

 \bibitem[Gratton et 
 al.(2013)]{Gr13} Gratton, R.~G., Lucatello, S., Sollima, A., et al.\ 2013, \aap, 549, A41 


\bibitem[Grevesse \& Sauval (1998)]{Gr98} Grevesse, N. \& Sauval, A.J. 1998,
  SSRv, 85, 161

 \bibitem[Geisler et al. (2007)]{Ge07} Geisler, D., Wallerstein, G., Smith,
 V.V., \&  Casetti-Dinescu, D.I. 2007, PASP, 119, 939

 \bibitem[Geisler et al. (2012)]{Ge12} Geisler, D., Villanova, S., Carraro, G., Pilachowski, C., 
 Cummings, J., Johnson, C. I., \& Bresolin, F. 2012, ApJ, 756, 40

 \bibitem[Harris (1996)]{Ha96} Harris, W.E. 1996, AJ, 112, 1487


 \bibitem[Johnson et al. (2006)]{Jo06} Johnson, J.A., Ivans, I.I., \& Stetson,
 P.B. 2006, ApJ, 640, 801

 \bibitem[Johnson \& Pilachowski (2010)]{Jo10} Johnson, C.I. \& Pilachowski, C.A. 2010, ApJ, 722, 1373
 
 \bibitem[Karakas \& Lattanzio (2007)]{Ka07} Karakas, A. \& Lattanzio, J.C.
 2007, PASA, 24, 103 

 \bibitem[Kock et al. (2009)]{Ko09} Koch, A., Cot\'e, P., \& McWilliam, A.
 2009, A\&A, 506, 729
 
 \bibitem[Kraft(1994)]{Kr94} Kraft, R.~P.\ 1994, \pasp, 106, 
 553


 \bibitem[Kurucz (1970)]{Ku70} Kurucz, R.L. SAO, 309

 \bibitem[Law \& Majewski (2010)]{La10} Law, D.R., \& Majewski, S.R. 2010,
 ApJ, 718, 1128

 \bibitem[Maccarone 
 \& Zurek(2012)]{Mac12} Maccarone, T.~J., \& Zurek, D.~R.\ 2012, \mnras, 423, 2 

 \bibitem[McWilliam(1997)]{Mc97} McWilliam, A.\ 1997, \araa, 35, 503 

 \bibitem[Mandushev et al. (1991)]{Ma91} Mandushev, G., Staneva, A., \&
 Spasova, N. 1991, A\&A, 252, 94

 \bibitem[Marigo et al. (2008)]{Mar08}

 \bibitem[Marino et al. (2008)]{Ma08} Marino, A.F., Villanova, S., 
 Piotto, G., Milone, A.P., Momany, Y., Bedin, L.R. \& Medling, A.M. 
 2008, A\&A, 490, 625

 \bibitem[Marino et al. (2009)]{Ma09} Marino, A.F., Milone, A.P., Piotto, G., 
 Villanova, S., Bedin, L.R., Bellini, A. \& Renzini, A. 2009, A\&A, 505, 1099
 
 \bibitem[Marino et al. (2011)]{Ma11} Marino, A.F., Villanova, S., Milone,
 A.P., Piotto, G., Lind, K., Geisler, D., \& Stetson, P.B. 2011, ApJ, 730L, 16


 \bibitem[Mashonkina et al. (2000)]{Ma00} Mashonkina, L.I.,  Shimanskii,
 V. V., \& Sakhibullin, N.A. 2000, ARep, 44, 790

 \bibitem[de Mink et 
 al.(2009)]{Mi09} de Mink, S.~E., Pols, O.~R., Langer, N., \& Izzard, R.~G.\ 2009, \aap, 507, L1 


 \bibitem[Monaco et al. (2005)]{Mo05} Monaco, L., Bellazzini, M., 
 Bonifacio, P., Ferraro, F.R., Marconi, G., Pancino, E., Sbordone, L., \&
 Zaggia, S. 2005, A\&A, 441, 141

 \bibitem[Mucciarelli et al. (2008)]{Mu08} Mucciarelli, A., Carretta, E., 
 Origlia, L., \& Ferraro, F.R. 2008, AJ, 136, 375

 \bibitem[Mucciarelli et al. (2009)]{Mu09} Mucciarelli, A., Origlia, L.,
 Ferraro, F.R.,\& Pancino, E. 2009, ApJ, 695, 134


 \bibitem[Reddy et al. (2003)]{Re03} Reddy, B.E., Tomkin, J., Lambert, D.L., \&
  Allende Prieto, C. 2003, MNRAS, 340, 304

 \bibitem[Reddy et al. (2006)]{Re06} Reddy, B.E., LambeSp05rt, D.L., \& Allende Prieto, C.
 2006, MNRAS, 367, 1329

 \bibitem[Roederer et al. (2009)]{Ro09} Roederer, I.U., Kratz, K.L., 
 Frebel, A., Christlieb, N., Pfeiffer, B., Cowan, J.J., \& Sneden, C.	
 2009, ApJ, 698, 1963

 \bibitem[Pignatari et al. (2008)]{Pi08} Pignatari, M., Gallino, R., 
 Meynet, G., Hirschi, R., Herwig, F., \& Wiescher, M. 2008, ApJ, 687, 95

 \bibitem[Piotto (2009)]{Pi09} Piotto, G. 2009, IAUS, 258, 233

 \bibitem[Piotto et al. (2012)]{Pi12} Piotto, G., Milone, A.P., 
 Anderson, J., Bedin, L.R., Bellini, A., Cassisi, S., Marino, A.F., Aparicio,
 A., Nascimbeni, V. 2012, ApJ, 760, 39

 \bibitem[Pompeia et al. (2008)]{Po08} Pomp\'eia, L., Hill, V., Spite, M., 
 Cole, A., Primas, F., Romaniello, M., Pasquini, L., Cioni, M.R., \& Smecker
 Hane, T. 2008, A\&A, 480, 379
 
 \bibitem[Sarajedini \& Layden (1997)]{Sa97} Sarajedini, A., \& Layden, A.
 1997, AJ, 113, 264

 \bibitem[Sbordone et al. (2007)]{Sb07} Sbordone, L., Bonifacio, P., Buonanno, R., Marconi, G.,
  Monaco, L, \&; Zaggia, S. 2007, A\&A, 465, 815

 \bibitem[Sbordone et al. (2011)]{Sb11} Sbordone, L., Salaris, M., Weiss, A.,
 \& Cassisi, S. 2011, A\&A, 534, 9

 \bibitem[Shetrone et al. (2001)]{Sh01} Shetrone, M.D., Cot\'e, P., \&
 Sargent, W.L.W. 2001, ApJ, 548, 592

 \bibitem[Sneden (1973)]{Sn73} Sneden, C. 1973, ApJ, 184, 839

 \bibitem[Sneden et al. (2008)]{Sn08} Sneden, C., Cowan, J.J., \& Gallino,
 R. 2008, ARA\&A, 46, 241


 \bibitem[Valcarce \& Catelan (2008)]{Va08} Valcarce, A. A. R., \& Catelan, M.
 2011, A\&A, 533, 120

 \bibitem[Ventura et al.(2001)]{Ve01} Ventura, P., D'Antona, 
 F., Mazzitelli, I., \& Gratton, R.\ 2001, \apjl, 550, L65 

 \bibitem[Villanova et al. (2010)]{Vi10} Villanova, S., Geisler, D., \&
 Piotto, G. 2010, ApJ, 722, 18



 \bibitem[Villanova \& Geisler (2011)]{Vi11} Villanova, S., \& Geisler, D.
 2011, A\&A, 535, 31

\end{thebibliography}
\end{document}